\documentclass[pra,twocolumn,superscriptaddress,nofootinbib]{revtex4}
\usepackage{amsmath}
\usepackage{amsfonts}
\usepackage{graphicx}

\begin{document}

\title{Degradation of a quantum reference frame}
\author{Stephen D. Bartlett}
\affiliation{School of Physics, The University of Sydney, Sydney, New South Wales 2006,
Australia}
\author{Terry Rudolph}
\affiliation{Optics Section, Blackett Laboratory, Imperial College London, London SW7
2BZ, United Kingdom}
\affiliation{Institute for Mathematical Sciences, Imperial College London, London SW7
2BW, United Kingdom}
\author{Robert W. Spekkens}
\affiliation{Department of Applied Mathematics and Theoretical
Physics, University of Cambridge, Cambridge CB3 0WA, U.K.}
\author{Peter S. Turner}
\affiliation{Institute for Quantum Information Science, University of Calgary, Alberta
T2N 1N4, Canada}
\date{19 April 2006}

\begin{abstract}
We investigate the degradation of reference frames, treated as
dynamical quantum systems, and quantify their \emph{longevity} as a
resource for performing tasks in quantum information processing. We
adopt an operational measure of a reference frame's longevity,
namely, the number of measurements that can be made against it with
a certain error tolerance.  We investigate two distinct types of
reference frame: a reference direction, realized by a spin-$j$
system, and a phase reference, realized by an oscillator mode with
bounded energy. For both cases, we show that our measure of
longevity increases quadratically with the size of the reference
system and is therefore non-additive.  For instance, the number of
measurements that a directional reference frame consisting of $N$
parallel spins can be put to use scales as $N^2$.
Our results quantify the extent to which microscopic or mesoscopic
reference frames may be used for repeated, high-precision
measurements, without needing to be reset -- a question that is
important for some implementations of quantum computing. We
illustrate our results using the proposed single-spin measurement
scheme of magnetic resonance force microscopy.
\end{abstract}

\maketitle

\section{Introduction}

In quantum measurement theory, the apparatus is generally treated as
a classical system -- one that is not described by the quantum
formalism.  The same is true for the macroscopic systems that serve
as reference frames, for instance, a ruler with respect to which
positions are defined, a set of gyroscopes with respect to which
orientations are defined, or a clock with respect to which phases
are defined.  Just as attempts to quantize the measurement apparatus
have led many researchers to foundational puzzles (such as the
quantum measurement problem), the quantization of reference frames
has also generated its fair share of confusion and controversy
(Ref.~\cite{BRS05} provides a synopsis and numerous references).

In addition to providing, through such puzzles, an opportunity for
us to refine our understanding of quantum theory, the quantization
of reference frames is also useful for answering certain practical
questions. In many quantum experiments, one can understand certain
systems as constituting mesoscopic or even microscopic reference
frames to which other systems are compared.  In such cases, the
conventional approach wherein reference frames suffer no back-action
may yield a poor approximation to a full quantum treatment.

A recent example is the investigation of how well a quantum optical
field prepared in a coherent state can serve as a local oscillator
for homodyne detection~\cite{Tyc04}. It has also been proposed that
the finite size of a reference frame can lead to effective
decoherence on the system it describes in contexts ranging from
collapse theories~\cite{Fin05} to the evaporation of black
holes~\cite{Gam04}. In particular, there have been several
investigations into the consequences of the quantum nature of laser
fields for the manipulation of physical qubits in quantum
information processing~\cite{Enk02,Gea02}.

Our goal in this paper is to go beyond consideration of how the
finite size of the reference frame affects the purity or coherence
of systems described with respect to it, and to examine how this
finiteness affects its \emph{longevity} i.e., how many times it can
be used in a measurement as an accurate reference frame.  In
particular, we demonstrate quantitatively how, if one makes
measurements of the ``orientation'' of a large number of systems
relative to a single quantum reference frame (RF), then as a result
of these measurements the state of the RF becomes more mixed, more
symmetric and is thereafter less useful for implementing the sorts
of tasks for which an RF is required. This \emph{degradation} (and
ultimate depletion) is yet another reason (in addition to those
provided in~\cite{BRS03,Vac03,BRS04a,Enk05}) for considering RFs as
an information-theoretic resource.

At first glance, one might expect that the longevity of an RF,
quantified by the number of estimations of the relative orientation
of a system to the RF that one can achieve with a certain error
bound before it degrades beyond use, would scale linearly with the
size of the RF.
For example, if two identical directional RFs were used together,
one would naively expect the resulting combined RF to last twice as
long as either constituent RF would separately.  However, our
analysis shows that the scaling does not match this expectation. In
fact, the longevity of such a directional RF scales
\emph{quadratically} with the number of spins $N$.  We show that the
same scaling holds for a phase reference, such as a coherent state.
Specifically, the longevity of such a phase reference scales
quadratically in the average excitation number (e.g., photon
number).  This result is encouraging for the potential usefulness of
mesoscopic reference frames in quantum information processing
applications.

We illustrate this point with a simple calculation for single-spin
measurements using magnetic resonance force microscopy,
demonstrating both that this measurement scheme may exhibit effects
due to the degradation of its directional RF with repeated use, but
also that it may be possible to use such a device for the large
number of measurement required for quantum computing applications
without significant degradation.

We emphasize the distinction between the \emph{quality} of an RF,
which describes how well a quantum RF approximates an ideal
classical RF for the purposes of measurement, and the
\emph{longevity} of an RF, which describes how many times it can be
used in a measurement whilst maintaining a certain quality.  Whereas
general uncertainty-principle-based arguments can often be used to
determine the scaling of the quality of a quantum RF, we are unaware
of any such argument that can predict the quadratic scaling of the
longevity that we determine in this paper.

\section{Preliminaries}

\label{subsec:Prelims}

We define the \emph{longevity} of a reference frame as the number of
times it can be used to perform a particular operational task with
some chosen finite degree of success. To explain the particular task
we choose, it is convenient to be able to compare our RF with
another, much larger ``background'' RF.  We denote the RF whose
longevity we are studying by $R$, which is initially correlated
(aligned) with the background RF.  (In Ref.~\cite{BRS05}, a RF $R$
with this feature was referred to as \emph{implicated}.)  For
example, $R$ may be a gyroscope used in a laboratory experiment, to
which directional systems are compared, and the background RF could
be the frame defined by the earth or the fixed stars.

The task to which $R$ will be put to use is the estimation of the
direction of a system $S$ relative to the background RF.  (We here
use the term ``direction'' in a generic way to mean the group
element relating $S$ to the background RF even if the the group of
transformations in which we are interested is not the rotation
group).  Such an estimation is achieved by measuring the relation
between $S$ and $R$, then combining the outcome with one's prior
information about the directionality of $R$ relative to the
background RF, to deduce something about the direction of $S$
relative to the background RF.  However, because $R$ is a finite
quantum system, the measurement of the relation between $S$ and $R$
causes a disturbance to the quantum state of $R$, so that after this
measurement, one's ability to infer the directionality of another
system $S'$ is decreased.  Thus, $R$ can only be used a finite
number of times for such a task before its directionality relative
to the background RF is so poorly defined that it is no longer
useful for this purpose.  It is this degradation that we
investigate.

Because neither $S$ nor $R$ physically interacts with the background
RF during the task of interest, we can take the background RF to be
non-dynamical. (More precisely, because the dynamics induced between
$R$ and $S$ is completely relational, we will see that the use of
the background RF is equivalent to simply choosing a convenient
gauge and has no bearing on the physics). The RF $R$, on the other
hand, must be treated dynamically.  Consequently, we assign a
Hilbert space and quantum states to $R$ but not to the background
RF. Adopting the terminology of Ref.~\cite{BRS05}, $R$ is treated as
an \emph{internal} RF, while the background RF is treated as an
\emph{external} RF.  We shall also say that we are treating $R$ as a
\emph{quantum} RF, and the background RF\ as a \emph{classical} RF.

A few aspects of our direction-estimation task must be made
specific. First, there is the nature of the measurement that
determines the relation between $S$ and $R$.  To be conservative, we
assume: (1) the measurement is the one that maximizes one's ability
to estimate the relation between $S$ and $R$, and thus also
maximizes one's ability to estimate the direction of $S$ relative to
the background RF (the figure of merit for estimation will be
specified later) and (2) the measurement is implemented in a manner
that leads to the smallest possible degradation of $R$ while still
satisfying feature (1). Feature (2) ensures that we are determining
the best possible longevity for a given degree of success in
direction-estimation. Taking into account the fact that the more
information we gain in the measurement, the more disturbance we
create, feature (1) ensures that the longevity we derive will be a
lower bound for the number of uses to which $R$ can be put for any
other sort of direction estimation task on $S$.


The second aspect of our estimation task that must be made specific
is the initial quantum state of $R$.  We choose to investigate the
state that leads to the least \emph{initial} error in estimating the
directionality of $S$ relative to the background RF.  We emphasize
that, although the optimal RF states and optimal measurements may
not necessarily be easy to achieve in practice, they establish the
quantum limit and therefore bound any RF longevity.

We note that an equivalent definition of longevity can be achieved
without making any reference to a background RF.  In this case, our
description of $S$ and $R$ would be subject to an effective
\emph{superselection rule}~for the group that is associated with the
RF \cite{BDSW04}. For instance, if the RF is for orientation, then
moving to a description that makes no reference to the background RF
would imply adopting an effective non-Abelian superselection rule
for the group of rotations SU(2).  Although the usefulness of this
mode of description has been emphasized elsewhere \cite{BRS05}, we
shall not make use of it in the present paper. In what follows, we
assign quantum states to $S$ and $R$ that are non-invariant under
the group of interest, even though the measurement of the relation
of $S$ to $R$ is invariant under this group. That is, the quantum
states are assigned relative to an arbitrary background RF, which is
not used in the measurement. One may liken this procedure to
choosing a gauge for convenience when describing gauge-invariant
processes.

\section{Directional reference frame}

In this section, we study the degradation of a directional RF,
that is, a reference for a direction in space. (Note that a
directional RF does not provide a full Cartesian frame, as it does
not provide a reference for rotations about its axis.)

\subsection{Measurement of a spin-1/2 system relative to a spin-$j$ directional RF}

\label{subsec:measurement}

We use a spin-$j$ system for our quantum directional RF, with
Hilbert space $\mathbb{H}_{j}$. The initial quantum state of the
spin-$j$ system is denoted $\rho ^{(0)}$. (Later, in
Sec.~\ref{sec:Performance}, we will determine the state that
minimizes the error in the estimation of direction.)  We describe
this RF relative to a background RF, as noted in
Sec.~\ref{subsec:Prelims}, and choose it to be aligned in the $+z$
direction relative to the background RF; we emphasise, though, that
this alignment with a background RF is essentially a choice of
gauge. Because the quantum RF will serve only as a reference
direction, and not a full frame, we can choose it to be invariant
under rotations about the $\hat{z}$ axis without loss of generality.
Thus, the initial quantum state, $\rho ^{(0)}$ is diagonal in the
$|j,m\rangle $ basis of $\mathbb{H}_{j}$, that is, the basis of
simultaneous eigenstates of $\hat{J}^2$ and $\hat{J}_z$.

The systems to be measured against the quantum RF will be spin-1/2
systems, each with a Hilbert space $\mathbb{H}_{1/2}$. We choose the
initial state of each such system to be the completely mixed state
$I/2$, where $I$ is the identity operator on $\mathbb{H}_{1/2}$.
This corresponds to maximal ignorance about the spin-1/2 system. Our
quantum RF will be used to measure many such independent spin-1/2
systems sequentially. We shall assume trivial dynamics between
measurements, and thus our time index will simply be an integer
specifying the number of measurements that have taken place. The
state of the RF following the $n$th measurement is denoted $\rho
^{(n)}$, with $\rho ^{(0)}$ denoting the initial state of the RF
prior to any measurement. We consider the state of the RF from the
perspective of someone who has \emph{not} kept a record of the
outcome of previous measurements.  Thus, at every measurement, we
average over the possible outcomes with their respective weights to
obtain the final density operator.

A measurement of the relative orientation of a spin-1/2 particle
to a spin-$j $ system is represented by operators that are
invariant under collective rotations. The measurement that
provides the maximum information gain about the relative
orientation between a spin-$j$ and a spin-$j^{\prime }$ system has
been determined in~\cite{BRS04b}. It is simply a measurement of
the magnitude of the total angular momentum $\hat{J}^{2}$. That
is, it is the projective measurement $\{\Pi _{J}\}$, where $\Pi
_{J}$ projects $\mathbb{H}_{j}\otimes \mathbb{H}_{j^{\prime }}$
onto the total $\hat{J}^{2}$ eigenspace with eigenvalue $J(J+1)$.
For the case where $j^{\prime }=1/2$, the optimal measurement for
determining the relative orientation is represented by the
two-outcome projective measurement $\{\Pi _{+}\equiv \Pi
_{j+1/2},\Pi _{-}\equiv \Pi _{j-1/2}\}$ on $\mathbb{H}_{j}\otimes
\mathbb{H}_{1/2}$~\cite{BRS04b}.

\subsection{Measurement-induced update of the directional RF state}

A given projective measurement can be associated with many
different update maps.  As discussed in Sec.~\ref{subsec:Prelims},
we choose the update map that is minimally disturbing.  In
App.~\ref{mindisturbingmaps}, it is shown that this corresponds to
adopting the \emph{L\"{u}ders rule} for updating: for a
measurement outcome associated with the projector $\Pi,$ the
quantum state $\rho $ is updated to $\Pi\rho \Pi/\mathrm{Tr}(\rho
\Pi).$

Thus, under the optimal measurement for relative orientation, the evolution
of the quantum RF as a result of the $n$th measurement is
\begin{equation}
  \rho ^{(n+1)}=\mathcal{E}(\rho ^{(n)})  \label{eq:DecohMap}
\end{equation}
where
\begin{equation}
  \mathcal{E}(\rho )=\text{Tr}_{S}\Bigl(\sum_{c\in \{+,-\}}\Pi _{c}(\rho
  \otimes I/2)\Pi _{c}\Bigr)\,,  \label{eq:DecohMap2}
\end{equation}
and $\text{Tr}_{S}$ denotes the partial trace over the spin-1/2 system.

The map $\mathcal{E}$ can be written using the operator-sum
representation~\cite{Nie00} as
\begin{equation}
  \mathcal{E}(\rho )=\frac{1}{2}\sum_{c\in \{+,-\}}\sum_{a,b\in
  \{0,1\}}E_{ab}^{c}\rho E_{ab}^{c\dag }\,,  \label{eq:RFdecoherence}
\end{equation}
where
\begin{equation}
  E_{ab}^{c}\equiv \left\langle a\right\vert \Pi _{c}\left\vert b\right\rangle
    \label{eq:EOps}
\end{equation}
is a Kraus operator on $\mathbb{H}_{j}$ and $\{\left\vert
0\right\rangle ,\left\vert 1\right\rangle \}$ is a basis for
$\mathbb{H}_{1/2}$. These operators can be straightforwardly
determined in terms of Clebsch-Gordon coefficients.

We note that the evolution of the RF is not unitary (and would not be
unitary even if the measurement result was kept), because the quantum RF
becomes entangled with the measured system which is subsequently discarded.
Thus, the map $\mathcal{E}$ is irreversible and consequently one may say
that the quantum state of the RF undergoes decoherence.

In App.~\ref{recurrenceforSU2}, we provide a recurrence relation for
the matrix elements of the state $\rho ^{(n)}$.

\subsection{Measure of directional RF quality}

\label{sec:Performance}

One simple measure of how much the RF degrades is the
fidelity~\cite{Nie00} between the state after $n$ measurements,
$\rho ^{(n)}$, and the initial state, $\rho ^{(0)}$, which clearly
decreases with $n$. Another natural measure is the
\emph{asymmetry} of the state~\cite{Vac05}, which also is found to
decrease with $n$. However, rather than use these measures, we
will instead choose to quantify the quality of the RF
operationally as the average probability of a successful
estimation of the orientation of a spin-1/2 system in a pure state
(relative to the background RF). For simplicity, we assume that
this fictional ``test'' spin-1/2 system is with equal probability
either aligned or anti-aligned with the background RF.  It is
worth making special note of the fact that in the sequence of
measurements that cause the RF degradation, the state of the
spin-1/2 system is \emph{not} assumed to be a pure state that is
aligned or anti-aligned with the RF.  It is assumed to be
completely mixed. However, after each measurement we ask: if a
spin-1/2 system in such a pure state \emph{were} compared to the
RF, how well could we estimate whether it was aligned or
antialigned?

Denote the pure state of the test spin-1/2 system that is aligned
(anti-aligned) with the initial RF by $|0\rangle$ ($|1\rangle$).
For a spin-1/2 system prepared in the state $|0\rangle$, the
probability of success, that is, of finding the correct ($j+1/2$)
result, is
\begin{equation}
P_{\mathrm{s}}^{0}(n) = \mathrm{Tr}\Bigl(\bigl[\rho^{(n)} \otimes |0\rangle
\langle 0|\bigr]\Pi _{+}\Bigr) =\mathrm{Tr}_{R}(\rho^{(n)} E_{00}^{+})\,.
\end{equation}
Similarly, for a spin-1/2 system prepared in the state $|1\rangle$,
the probability of the correct $(j-1/2)$ result is
$P_{\mathrm{s}}^{1}(n)=\mathrm{Tr}_{R}(\rho^{(n)} E_{11}^{-})$.
Thus, the average probability of success is
\begin{equation}
\overline{P}_{\mathrm{s}}(n)=\frac{1}{2}\mathrm{Tr}_{R}(\rho^{(n)}
(E_{00}^{+}+E_{11}^{-}))\,.
\end{equation}
Using Clebsch-Gordon coefficients to determine the Kraus operators
of Eq.~(\ref{eq:EOps}), we find
\begin{equation}
  \label{eq:Psuccess}
  \overline{P}_{\text{s}}(n) =1-\tfrac{1}{(2j+1)}\sum_{m=-j}^{j}(j-m+1/2)
\textrm{Tr}_R(\rho^{(n)} |j,m\rangle \langle j,m|) \,.
\end{equation}
We remind the reader that, at every measurement, the resulting
density matrix for the RF has been averaged over the possible
previous outcomes with their respective weights.  Thus, this measure
$\overline{P}_{\text{s}}(n)$ is applicable to situations where the
prior measurement record is discarded.  In addition, this method
also quantifies the \emph{average} case longevity even if the
measurement record is kept.

We now focus our attention on a particular initial state of the RF.
We choose the state that is optimal for our measure of RF quality,
in other words, we choose the $\rho ^{(0)}$ that maximizes
$\overline{P}_{\text{s}}(0)$.  This involves minimizing the sum in
Eq.~(\ref{eq:Psuccess}), which may be written as
\begin{equation}
  \mathrm{Tr}(M\rho ^{(0)})\,,
\end{equation}
where
\begin{equation}
  M=\sum_{m=-j}^{j}(j-m+1/2)\left\vert j,m\right\rangle
  \left\langle j,m\right\vert\,.
\end{equation}
Clearly then, we should choose $\rho ^{(0)}$ to be an eigenvector
of $M$ associated with the smallest eigenvalue, namely, $\rho
^{(0)}=\left\vert j,m=j\right\rangle \left\langle
j,m=j\right\vert.$ Thus we see that the quantum state that
maximizes our measure of quality of the RF is an SU(2) coherent
state, yielding $\overline{P}_{\mathrm{s} }(0)=1-1/[2(2j+1)]$.


(This result is not in conflict with that of Peres and
Scudo~\cite{Per01}, where it was demonstrated that the product
state of $N$ parallel spins (which is equivalent to the SU(2)
coherent state) was \emph{not} the optimal state for the
transmission of a reference direction, because their system was
not assumed to be a spin-$j$ representation, as we have done here.
An interesting direction for future work is to extend the results
of~\cite{Lin05} in order to determine the optimal measurements of
spin-1/2 systems against these more general RF systems and to
explore the degradation in this case.)

\subsection{Scaling of the longevity of a directional RF with size}

Our problem is to solve Eq.~(\ref{eq:DecohMap}) to determine the
RF state as a function of $n$. In the limit $j\gg 1,$ one can
obtain a closed-form expression for the diagonal elements of the
state of the RF at step $n$, assuming an initial state $\rho
^{(0)}=|j,j\rangle \langle j,j|$. It is
\begin{multline}
\textrm{Tr}(\rho^{(n)} |j,m\rangle \langle j,m|)
\\
=(-1)^{j-m}\binom{n}{j-m}(j-m)!(-2j)U(j-m+1,n+2,-2j)\,,
\label{DiagMatrixElementsLimit}
\end{multline}
where $U(a,b,z)$ is Kummer's $U$ function~\cite{AS}. Details are
given in App.~\ref{recurrenceforSU2}.

\begin{figure}
\begin{center}\includegraphics[width=9cm]{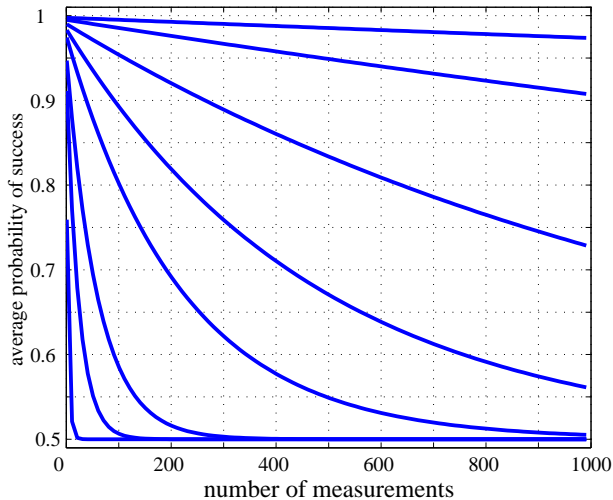}\end{center}
\caption{A plot of the success probability for a test particle to be
measured correctly against the directional reference frame, versus
the number of previous spins that have been measured against the
frame. The plots (from bottom left to top right) are for a reference
frame of spin $j=1, 3, 5, 10, 15, 25, 50, 100$, respectively.
\label{fig:ProbSuccess} }
\end{figure}

Using this result, we find the rate at which
$\overline{P}_{\text{s}}(n)$ decreases with $n$ to be
\begin{equation}
  \overline{P}_{\text{s}}(n)=\tfrac{1}{2}+\tfrac{j}{2j+1}(1-\tfrac{2}{(2j+1)^{2}}
  )^{n}\,.
  \label{eq:PsuccessSU(2)CS}
\end{equation}
We plot this for various values of $j$ in Fig.~\ref{fig:ProbSuccess}.

The initial slope of this function bounds the rate of degradation. It is
\begin{equation}
  R \equiv \overline{P}_{\text{s}}(1)-\overline{P}_{\text{s}}(0) =-2j/(2j+1)^{3}\,.
\end{equation}
Thus, in the large $j$ limit, we have the rate of degradation with $n$
satisfying $R\geq -1/(4j^{2})$.

Let $\epsilon <1$ be a fixed allowed error probability for the
spin-1/2 direction estimation problem. After $n$ measurements, the
probability of successful estimation is lower bounded by $1+nR$, so
the number of measurements required to ensure that this bound be
greater than $1-\epsilon$ is $-\epsilon/R$. Consequently, the number
of measurements that can be implemented relative to the spin-$j$ RF
with probability of error less than $\epsilon$ is
\begin{equation}
  n_{\mathrm{max}}\simeq \epsilon j^{2}\,.
\end{equation}
This result implies that the number of measurements for which an
RF is useful, that is, the longevity of a RF, increases
\emph{quadratically} rather than linearly with the size of the RF.

\begin{figure}
\begin{center}\includegraphics[width=9cm]{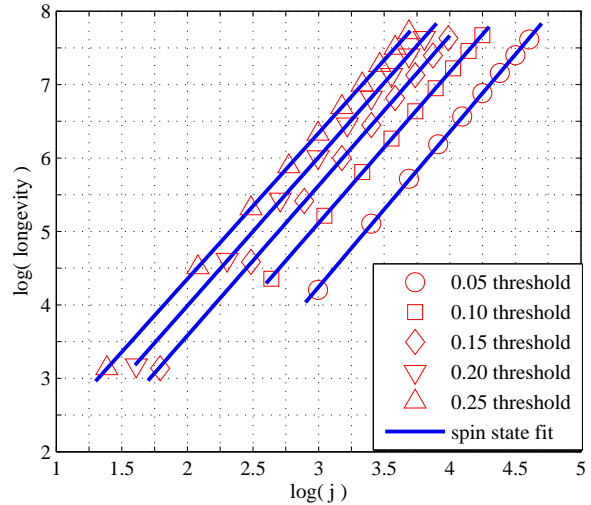}\end{center}
\caption{ \label{fig:spinlog2plot} A loglog plot of the longevity
(number of measurements before the error probability for a test
particle exceeds a fixed threshold) of a directional reference frame
versus the size of the frame. Plots for various different error
thresholds are given. A clear quadratic asymptotic dependence
emerges despite the relatively small reference frames indicated:
linear regression yields gradients which all lie within $2\pm 0.1$.}
\end{figure}

Numerical calculation of the longevity's dependence on RF size for
various choices of error threshold confirm this result; see
Fig.~\ref{fig:spinlog2plot}.

This result implies that, in order to maximize the number of
measurements that can be achieved with a given error thereshold, one
should combine all of one's directional RF resources into a single
large RF and perform all measurements relative to it.  It is also
worth noting that if the directional RF is composed out of $N$
spin-1/2 systems, so that it has size $j=N/2+1,$ the number of
measurements that can be performed with probability of error less
than $\epsilon $ scales as $N^{2},$ whereas one might have
na\"{\i}vely thought that it would be good for only as many
measurements as there are constituent spins.  Finally, note that an
SU(2) coherent state for $N$ spin-1/2 systems is a product state so
that entanglement does not appear to be responsible for this
property.

\subsection{Application to single-spin measurement using magnetic resonance
force microscopy for solid-state quantum computing}

The physical nature of measurement apparatuses and reference
systems is particularly relevant for quantum computation. The
stringent requirements placed on measurement devices for use in
quantum computing include (i) large coupling strengths, often
requiring the measurement device to be made very small and placed
close to the quantum registers (qubits); and (ii) the apparatuses
must be well-described by classical, noiseless devices in order to
obtain the required levels of accuracy (e.g., perform projective
von Neumann measurements). These two requirements appear at first
glance to be mutually exclusive, but may be satisfied by novel
mesoscopic measurement schemes. Consider, as an example, the
proposal to use magnetic resonance force microscopy (MRFM) to
measure the direction of a single spin~\cite{Sid95,Rug04} in a
solid-state quantum computer. This scheme makes use of a small
magnet placed on the tip of a nano-mechanical resonator;
monitoring the oscillations of the resonator can be used to
measure the direction of a spin placed near the magnet. To couple
strongly to a single spin, the measurement device must be cold and
very small (i.e., a resonator with a mass of the order of
picograms, with a magnet on the tip of length scale 10
nm~\cite{Sid95}). However, objects on this scale are poorly
approximated as classical, macroscopic measurements apparatuses,
and thus a semi-classical or full quantum treatment of the
measuring device is required.

The results we have presented here can be applied to the proposed use of
MRFM to measure the direction of a single spin, and its associated
applications to solid-state quantum computing. 
The design parameters of a scheme to obtain single-spin measurement
are outlined in~\cite{Sid95}, in which the magnet consists of an
iron sphere of diameter $\sim 300$ \AA .  Treated as a quantum
system, then, the magnet can be approximately described by $10^6$
parallel spins. With an error tolerance of $10^{-4}$ (a reasonable
threshold for quantum computation) this RF would perform $10^8$
optimal measurements before degrading beyond further usefulness.
Thus, if the measurement scheme can be chosen such that the
disturbance on the RF due to each measurement is comparable with
that of the optimal update map, it appears that a mesoscale RF such
as this may be suitable for repeated use in large-scale quantum
computation.

\section{Phase reference}

We now consider the case of a phase reference. Our systems will
consist of oscillator modes, taken to be optical modes for
clarity.\ It should be noted that we focus on the optical case
only to make the description of the results less abstract; the
results are applicable to any phase reference.

\subsection{Measurement of single-rail qubit relative to phase
reference}

For our phase reference, we take a single oscillator mode with
Hilbert space denoted by $\mathbb{H}_{R}$. (Note that we could
instead use a multimode phase reference, or even many qubits
prepared as ``refbits''~\cite{Enk05}.)  Again, the initial quantum
state of the RF is denoted $\rho ^{(0)}$.  We describe this RF
relative to a background phase reference, and arbitrarily choose
its phase to be zero (our choice of gauge).

The systems to be measured against the RF will be single-rail
qubits, i.e., oscillators with states restricted to the
lowest-energy 2-dimensional subspace spanned by Fock states
$|0\rangle $ and $|1\rangle$. The Hilbert space is denoted
$\mathbb{H}_{S}$.

We consider the particular estimation task wherein the system is
promised to encode phase $0$ or phase $\pi $ with equal probability,
which is to say that it is promised to be in a state $\left\vert
+\right\rangle$ or $\left\vert -\right\rangle$, where $\left\vert
\pm \right\rangle =2^{-1/2}\left( \left\vert 0\right\rangle \pm
\left\vert 1\right\rangle \right)$, with equal probability.  In this
case, an optimal measurement for estimating whether the relative
phase between system and quantum RF is $0$ or $\pi$ is the
two-outcome projective measurement $\{\Pi _{+},\Pi _{-}\}$ with
\begin{align}
  \Pi _{+}& =\sum_{m=1}^{\infty }|m,{+}\rangle \langle m,{+}|
  +|0\rangle _{R}\left\langle 0\right\vert \otimes \left\vert 0\right\rangle
  _{S}\langle 0|  \nonumber \\
  \Pi _{-}& =\sum_{m=1}^{\infty }|m,{-}\rangle \langle m,{-}|\,,
  \label{Pi+Pi-}
\end{align}
where
\begin{equation}
  |m,\pm \rangle =\frac{1}{\sqrt{2}}\bigl(|m\rangle _{R}|0\rangle
  _{S}\pm |m{-} 1\rangle _{R}|1\rangle _{S}\bigl)\,.
  \label{mpm}
\end{equation}
This is demonstrated in App.~\ref{optimalphasemmt}.

It is worth noting that this measurement is a coarse-graining of
the optimal projective measurement of phase~\cite{Bag05} on each
of the 2-dimensional eigenspaces of total photon number; we
further note that this measurement was used in~\cite{Ver03} for
the same purpose.

We will be interested in determining how the RF state evolves
after many such measurements on distinct systems. For simplicity,
we consider the RF state as it is described by an observer who
does not know whether the system was in the state $\left\vert
+\right\rangle $ or $\left\vert -\right\rangle $ initially, and
because the two are presumed equally likely, the initial state of
the system is presumed to be the completely mixed state $I$/2. We
also assume that the observer does not keep track of the results
of the measurements.

\subsection{Measurement-induced update of the phase reference}

As with the directional example, the update rule for this
two-outcome projective measurement that is minimally-disturbing is
the L\"{u}ders rule, described in App.~\ref{mindisturbingmaps}.

With this choice of update rule, the evolution of the quantum RF
as a result of the $n$th measurement has the same form as
Eqs.~(\ref{eq:DecohMap}-\ref{eq:EOps}) but where the $E_{ab}^{c}$
operators are now given by
\begin{gather}
  E_{00}^{+}=\tfrac{1}{2}(I_{R}+|0\rangle \langle 0|)\,,\quad E_{00}^{-}=
  \tfrac{1}{2}(I_{R}-|0\rangle \langle 0|)\,, \\
  E_{11}^{+}=E_{11}^{-}=\tfrac{1}{2}I_{R}\,, \\
  E_{10}^{+}=-E_{10}^{-}=(E_{01}^{+})^{\dag }=-(E_{01}^{-})^{\dag }=\tfrac{1}{2}A\,,
\end{gather}
where $I_{R}$ is the identity on $\mathbb{H}_{R}$ and
\begin{equation}
  A\equiv \sum_{m=0}^{\infty }|m\rangle \langle m+1|\,.
  \label{A}
\end{equation}
The update map due to a single measurement is therefore
\begin{equation}
  \mathcal{E}(\rho )=\tfrac{1}{2}\rho +\tfrac{1}{4}|0\rangle \langle 0|\rho
  |0\rangle \langle 0|+\tfrac{1}{4}A^{\dag }\rho A+\tfrac{1}{4}A\rho A^{\dag
  }\,.  \label{eq:RFdecoherenceA}
\end{equation}

Again we note that although it may be difficult to achieve such an
update map in practice, it is nonetheless interesting to consider
because it defines the quantum limit of RF longevity. If the
systems under consideration are modes of the electromagnetic
field, this update map is particularly impractical given current
technology because it corresponds to a quantum non-demolition
measurement on the electromagnetic field. An interesting problem
for future research is to determine how much degradation, in
excess of the quantum limit, is incurred by the use of more
realistic update maps.

\subsection{Measure of phase reference quality}

As with the directional case, we operationally quantify the quality
of the RF at step $n$ as the average probability of success in a
hypothetical measurement of whether a qubit system is in phase or
out of phase with the background phase reference, that is, whether
the qubit is in the state $\left\vert +\right\rangle $ or
$\left\vert -\right\rangle ,$ given equal prior probability for
these two possibilities.  Denoting the state of the RF at step $n$
by $\rho ^{(n)}$, and using the optimal measurement~(\ref{Pi+Pi-}),
the average success rate is found to be
\begin{equation}
  \overline{P}_{\mathrm{s}}(n)=\frac{1}{2}
  +\frac{1}{2}\sum_{m=0}^{\infty}\text{Re}\Bigl(\rho _{m,m+1}^{(n)}\Bigr)\,.
  \label{eq:Psuccessphase}
\end{equation}
where $\rho_{m,m+1}^{(n)}\equiv \textrm{Tr}(\rho^{(n)} |m\rangle
\langle m+1|)$.

Just as we did for the directional case, we consider the degradation
in the case of an initial state of the RF that is initially optimal
for our measure of RF quality, i.e., the state $\rho ^{(0)}$ that
maximizes $\overline{P}_{\text{s}}(0)$.  However, if the accessible
Hilbert space $\mathbb{H}_R$ is taken to be infinite-dimensional,
then this probability can trivially be made arbitrarily close to
one. To obtain a non-trivial and physical result, we must bound the
energy of the state $\rho^{(0)}$ in some way; we may then use this
bound as a measure of the ``size'' of the RF, analogous to the use
of total spin $j$ for the directional case.

One natural choice for implementing this bound is to limit the
maximum photon number to some fixed value $N$, i.e., demand that
$\rho^{(0)}$ has support entirely on the $(N+1)$-dimensional
subspace spanned by $\{|m\rangle\,,\ m=0,\ldots,N \}$.  We now
seek the state $\rho^{(0)}$ that maximizes
$\overline{P}_{\text{s}}(0)$ subject to this constraint. This
requires us to maximize the expression
\begin{equation}
  \mathrm{Tr}(M_{N}\rho ^{(0)})\,,
\end{equation}
where
\begin{equation}
  M_{N}=\sum_{m=0}^{N}(|m\rangle \langle m+1|+|m+1\rangle \langle m|)\,.
\end{equation}
Thus we must choose $\rho ^{(0)}$ to be the eigenvector of $M_{N}$
with minimal eigenvalue.  The solution is
\begin{equation}
  \left\vert \psi _{N}\right\rangle =\mathcal{N}\sum_{m=0}^{N}\sin \Bigl[
  \frac{(m+1)\pi}{N+2} \Bigr] \left\vert m\right\rangle\, ,
  \label{eq:optimalphasestate}
\end{equation}
where $\mathcal{N}$ is a normalization factor.  This is the same
state that is optimal for determining the relative path length
between two arms in an interferometer, for several different figures
of merit~\cite{Sum90}. We provide a derivation of
Eq.~(\ref{eq:optimalphasestate}) in App.~\ref{optimalphasestate}.
For this state, we have
\begin{align}
  \overline{P}_{\mathrm{s}}(0)
  &=\frac{1}{2}\bigl(1+\cos(\pi/(N+2))\bigr) \nonumber \\
  &\simeq 1 - \frac{\pi^2}{2N^2}\,, \ \text{for large}\ N \,.
\end{align}

We note, however, that our method of bounding the energy of the
initial state $\rho^{(0)}$ excludes states, such as the coherent
state $|\alpha\rangle =
e^{-|\alpha|^2/2}\sum_{n=0}^{\infty}(\alpha^n/\sqrt{n!})|n\rangle$,
that have non-zero support on the entire Hilbert space
$\mathbb{H}_R$ but nevertheless have a finite energy.  It is beyond
the scope of this paper to explore all the varied possibilities for
determining optimal phase references for alternate choices of
bounds.  However, because of the ubiquity of coherent states as
phase references in many physical systems, we will also consider how
the longevity of a coherent state compares with the optimal
phase-encoding state with bounded
number~(\ref{eq:optimalphasestate}). We compare the size of such
states through \emph{average} photon number $\bar{n}$, which is
given by $|\alpha|^2$ for a coherent state and by $N/2$ for an
optimal phase-encoding state containing at most $N$ photons.

\subsection{Degradation of phase reference}

\begin{figure}
\begin{center}\includegraphics[width=9cm]{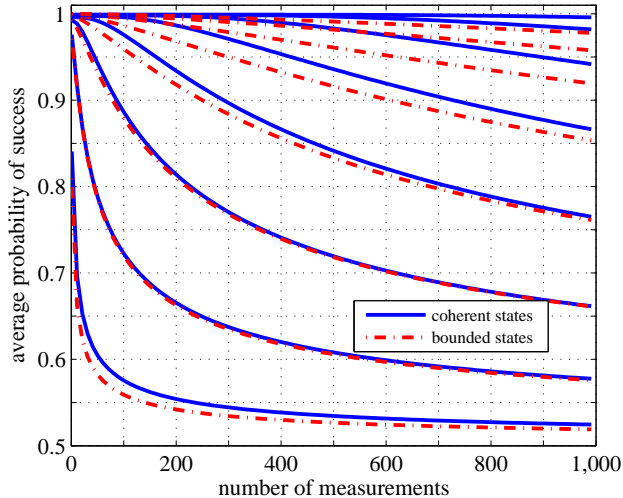}\end{center}
\caption{A plot of the success probability for a test mode to be
measured correctly against the phase reference frame, versus the
number of previous oscillators that have been measured against the
frame. Plots are shown for both coherent states and the optimal
phase-encoding states for bounded number. The plots (from bottom
left to top right) are for a reference frame of average photon
number $\bar{m}=1, 4, 9, 16, 25, 36, 49, 64$, respectively.  The
fact that the plots for the coherent state and the optimal
phase-encoding state are nearly overlapping at $\bar{m} = 4$ and $9$
is a result of the similarity of the number distributions of the two
states at these values of $\bar{m}$.\label{fig:cohoprplot2}}
\end{figure}

To determine the rate at which $\overline{P}_{\text{s}}(n)$
decreases with $n $, given an initial state $\rho^{(0)}$ for the
RF, we need only determine the evolution of the matrix elements
$\rho_{m,m+1}$. To this end, we use Eq.~(\ref{eq:RFdecoherenceA})
to obtain the following recurrence relation for these elements,
\begin{equation}
  \rho _{m,m+1}^{(n+1)}=\frac{1}{2}\rho _{m,m+1}^{(n)}
  +\frac{1}{4}\rho_{m+1,m+2}^{(n)}+\frac{1}{4}\rho _{m-1,m}^{(n)}\,.
\end{equation}
As it turns out, the elements on the $m,m+1$ diagonal depend only on
elements on this diagonal at the previous time step.  The solution
to this recurrence relation is found to be
\begin{equation}
  \rho _{m,m+1}^{(n)}=\frac{1}{4^{n}}\sum_{j=1}^{n+m+1}\Bigl[ \tbinom{2n}{n-j+m}
  -\tbinom{2n}{n-j-m}\Bigr] \rho _{j,j+1}^{(0)}\,.
\end{equation}
This expression can be used in Eq.~(\ref{eq:Psuccessphase}) to
determine numerically how $\overline{P}_{\text{s}}(n)$ decreases
with $n$ for a given choice of initial state $\rho^{(0)}$.  In
Fig.~\ref{fig:cohoprplot2}, we plot $\overline{P}_{\text{s}}(n)$ for
an initial optimal phase-encoding state with number bound $N$,
Eq.~(\ref{eq:optimalphasestate}), for various values of $N$. For
comparison, we also plot $\overline{P}_{\text{s}}(n)$ for an initial
coherent state with mean number $|\alpha|^2=N/2$ (so that we are
comparing states with the same mean number of photons).  We note
that, although the optimal phase-encoding state with bounded number
gives a superior \emph{initial} probability of success
$\overline{P}_{\text{s}}(0)$ when compared to that of the coherent
state, the latter appears more robust and does not degrade as
quickly.  This result demonstrates, perhaps surprisingly, that
optimizing the initial quality of a quantum reference frame
(quantified by the initial probability of success) does not optimize
the longevity of the state.  It is an interesting line of future
research to determine which quantum states give the optimal
longevity.

\begin{figure}
\begin{center}\includegraphics[width=9cm]{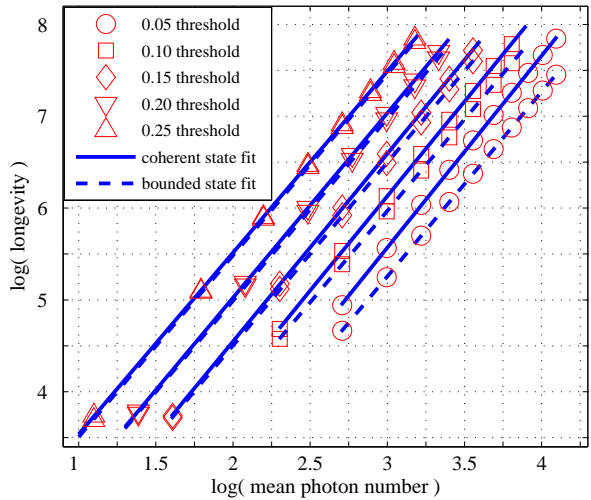}\end{center}
\caption{A loglog plot of the longevity (number of measurements
before the error probability for a test particle exceeds a fixed
threshold) of a phase reference frame versus the size of the frame.
Plots for various different error thresholds are given. A clear
quadratic asymptotic dependence emerges despite the relatively small
reference frames indicated: linear regression yields slope fits
which all lie within $2\pm 0.1$. \label{fig:cohoprlog2plot} }
\end{figure}

We find numerically that the longevity of a phase reference, both
for the optimal phase-encoding state with bounded number and for
the coherent state, scales as $\bar{n}^2$, the mean photon number
squared; see Fig.~\ref{fig:cohoprlog2plot}.  Specifically, we find
that $\overline{P}_{\text{s}}(n)$ reaches a fixed value
\emph{independent} of $\bar{n}$ after a number of uses equal to
$\bar{n}^2$.  Again, we find that the number of measurements for
which a phase RF is useful, that is, the longevity of a RF,
increases \emph{quadratically} rather than linearly with the size
of the RF.

\section{Discussion}

We have presented results on the degradation of both a directional
and a phase reference. In each case we have essentially uncovered a
quadratic scaling of the number of uses the frame can endure in
terms of the size of the frame.  An open question is whether this
quadratic scaling of longevity holds for all types of RFs.

It is interesting to consider whether a semi-classical model would
suffice for describing the degradation of an RF.  In particular, one
could employ a description of a dynamical classical RF that uses a
probability distribution on the space of ``orientations'' which
encodes one's ignorance of the exact direction, and also use a
measurement theory that relates the probability of successful
measurement to the uncertainty (variance) of this distribution. This
uncertainty would increase with repeated measurements; a simple
model might be to use diffusion (i.e., a classical random walk) to
characterise this increase.  Our key result in this paper is to
determine the \emph{rate} of degradation and its dependence on the
size of the RF.  If this rate was to be determined solely from a
semi-classical model, then the model would require a sophisticated
measurement theory for determining the direction of a quantum spin
relative to a classical object of a given size.  We believe that a
heuristic, semi-classical model that predicts the quadratic scaling
of the number of uses of an RF in terms of its size would be very
desirable.

It is worth noting that we have considered only the consequences of
measurements on an RF, and not the consequences of a continual
interaction with an environment.  In practice, RFs \emph{do}
interact with their environments, and so the impact on their
longevity is an important question for future research.  It should
be noted that although many environments are likely to act in a
manner similar to measurements, thereby increasing the rate of
degradation, others might in fact act to continuously realign one's
quantum RF with another, larger, RF which exists in the background.
For instance, if the RF is composed of a number of spins, and is
continuously interacting with a strong uniform magnetic field, then
thermalization will tend to cause the spins to become aligned with
the field.

Spatial/directional RFs are ubiquitous, and so one might question
the need for considering their degradation. It could be argued, for
example, that any directional RF which has degraded can be realigned
with some background RF in the lab (treated as a macroscopic
directional RF) at any time. However, such realignment may prove to
be difficult for microscopic apparatuses that are placed close to
the quantum systems to be measured.  In addition, realignment of a
degraded \emph{shared} RF costs resources - an issue that can be
important especially within the context of quantum communication.
Our main result shows, however, that such realignment may not be
necessary for many situations, such as the MRFM example discussed
above.

Although some RFs, such as directional RFs, are indeed ubiquitous,
RFs for many other degrees of freedom are not. Some RFs must be
painstakingly prepared via some controlled quantum process. One
example is the BCS ground state of a superconductor, which serves
as a phase reference for experiments involving superconducting
qubits; another example is a Bose-Einstein condensate. An analysis
similar to the one presented in this paper for any such frame
could be used to determine the RF size necessary to keep errors
below a specified threshold for a specified number of uses.

Although in this paper we have considered the amount of degradation
that results from measurements that maximize the information gain
about the relative orientation of system and RF, it is clear that
there is a trade-off between longevity and information gain; the
more informative the measurement, the more degradation to the RF. An
interesting problem for future research is to determine the precise
nature of this tradeoff.

\begin{acknowledgments}
The authors gratefully acknowledge Andrew Doherty for helpful
discussions, and the Perimeter Institute for support.  SDB
acknowledges support from the Australian Research Council.  TR
acknowledges support from the Engineering and Physical Sciences
Research Council of the United Kingdom.  RWS acknowledges support
from the Royal Society.  PST acknowledges support from the Alberta
Ingenuity Fund and the Informatics Circle of Research Excellence.
\end{acknowledgments}

\appendix

\section{Proof that L\"{u}ders rule updating is minimally-disturbing }

\label{mindisturbingmaps}

A measure of the quality of a RF, henceforth a measure of
``frameness'', is operational if it quantifies the maximum success
with which one can achieve some task in a variation over all
protocols that use only the resource and invariant operations.  Any
operational measure of frameness $F$ is by definition a monotone
with respect to operations that are invariant under the action of
the group associated with that RF, that is, if $\rho $ is the
quantum state of a RF for the group $G$, then
\begin{equation}
  F(\mathcal{E}(\rho ))\leq F(\rho )\,,
\end{equation}
for all $\mathcal{E}$ that are invariant under $G,$ that is, for
which $\mathcal{E}\left[ U(g)(\cdot )U^{\dag }(g)\right]
=U(g)\mathcal{E}(\cdot )U^{\dag }(g)$ for all $g\in G$.  We call
such measures \emph{frameness monotones}.

The maximum probability of a successful estimation of a direction of
a system relative to a quantum reference frame is clearly an
operational measure of frameness (it involves a variation over all
estimation protocols that use invariant operations and the RF
resource).  It is therefore by definition a frameness monotone.

We now make use of the following proposition concerning CP maps that
are update maps for a projective measurement.

\textbf{Proposition:} For any map $\mathcal{E}$ such that
$\mathcal{E}^{\dag }(I)=\Pi,$ where $\Pi$ is a projector, we have
\begin{equation}
  \mathcal{E}(\rho)=\mathcal{E}(\Pi\rho \Pi)\,.
\end{equation}
\textit{Proof.}  Any Krauss decomposition of $\mathcal{E}$ is of the
form
\begin{equation}
  \mathcal{E}(\rho)=\sum_{\mu}K_{\mu}\rho K_{\mu}^{\dag}\,,
\end{equation}
with
\begin{equation}
  \sum_{\mu}K_{\mu}^{\dag}K_{\mu}=\Pi\,.
\end{equation}
Given that every term in the sum is positive, that is,
$K_{\mu}^{\dag}K_{\mu }\geq 0$, it follows that
\begin{equation}
  \text{supp}(K_{\mu}^{\dag}K_{\mu})\subseteq\text{supp}(\Pi)\,,
\end{equation}
where supp$(A)$ denotes the support on the Hilbert space of the
operator $A$. \ It follows that
\begin{equation}
  \text{supp}(\sqrt{K_{\mu}^{\dag}K_{\mu}})\subseteq\text{supp}(\Pi)\,,
\end{equation}
and thus
\begin{equation}
  \sqrt{K_{\mu}^{\dag}K_{\mu}}=\sqrt{K_{\mu}^{\dag}K_{\mu}}\Pi\,.
\end{equation}
But now using the polar decomposition
$K_{\mu}=U_{\mu}\sqrt{K_{\mu}^{\dag }K_{\mu}}$, we infer that
\begin{equation}
  K_{\mu}=K_{\mu}\Pi\,.
\end{equation}
Our claim then follows trivially. QED.

Now, by virtue of the fact that $\mathcal{E}$ is invariant under
$G$, and by virtue of the fact that the probability of successful
estimation of a direction, $P_{\text{s}},$ is a frameness
monotone, it follows that
\begin{align}
  P_{\text{s}}(\mathcal{E}(\rho ))& =P_{\text{s}}(\mathcal{E}(\Pi\rho \Pi)) \\
  & \leq P_{\text{s}}(\Pi\rho \Pi)
\end{align}
Normalizing our density operators, we have
\begin{equation}
  P_{\text{s}}(\frac{\mathcal{E}(\rho )}{\text{Tr}\left( \Pi\rho \right) })\leq
  P_{\text{s}}(\frac{\Pi\rho \Pi}{\text{Tr}\left( \Pi\rho \right) })\,.
\end{equation}
Thus the probability of successful estimation given a L\"{u}ders
rule collapse map $\rho \rightarrow \Pi\rho
\Pi/\mathrm{Tr}(\Pi\rho )$ is an upper bound for the probability
of successful estimation for any collapse map associated with
$\Pi.$ \ \ Thus one can do no better than to use the L\"{u}ders
rule collapse map.

\section{Derivation of the recurrence relation for a directional RF}

\label{recurrenceforSU2}

In order to explicitly solve for the degraded state $\rho ^{(n)}$ of a
directional RF as a function of $n$, we now derive a recurrence relation for
the matrix elements of the state. The $E_{ab}^{c}$ operators can be obtained
in the $|j,m\rangle $ basis using Clebsch-Gordon coefficients by making use
of the decomposition
\begin{equation}
\Pi _{\pm }=\sum_{M=-(j\pm \frac{1}{2})}^{j\pm \frac{1}{2}}|j\pm
\tfrac{1}{2} ,M\rangle \langle j\pm \tfrac{1}{2},M|\,,
\end{equation}
where
\begin{multline}
|j\pm \tfrac{1}{2},M\rangle _{RS} \\
=(j,M-\tfrac{1}{2};\tfrac{1}{2},\tfrac{1}{2}|j\pm \tfrac{1}{2},M)
|j,M-\tfrac{1}{2}\rangle _{R}|0\rangle _{S} \\
+(j,M+\tfrac{1}{2};\tfrac{1}{2},-\tfrac{1}{2}|j\pm
\tfrac{1}{2},M)|j,M+ \tfrac{1}{2}\rangle _{R}|1\rangle _{S}\,.
\end{multline}
Denoting $\left\langle j,m\right\vert _{R}E\left\vert j,m^{\prime
}\right\rangle _{R}$ by $[E]_{xx^{\prime }},$ where $x\equiv j-m$, we have
\begin{equation*}
\begin{array}{rclcrcl}
\big[E_{00}^{+}\big]_{xy} & = & \frac{2j+1-x}{2j+1}\delta _{xy}, &
\big[
E_{11}^{+}\big]_{xy} & = & \frac{(x+1)}{2j+1}\delta _{xy} &  \\
\big[E_{00}^{-}\big]_{xy} & = & \frac{x}{2j+1}\delta _{xy}, &
\big[E_{11}^{-}
\big]_{xy} & = & \frac{2j-x}{2j+1}\delta _{xy} &  \\
\big[E_{01}^{\pm }\big]_{xy} & = &
\multicolumn{5}{c}{\big[E_{10}^{\pm }\big] _{yx}=\pm
\frac{\sqrt{x(2j+1-x)}}{2j+1}\delta _{x-1,y},}
\end{array}
\end{equation*}
where $0\leq x,y\leq 2j$.

Using these expressions, we are led to the following recurrence
relation:
\begin{multline}
(2j+1)^{2}\rho _{xy}^{(n+1)}=(2j^{2}+2j+1+2(j-x)(j-y))\rho _{xy}^{(n)} \\
+\sqrt{xy(2j+1-x)(2j+1-y)}\rho _{x-1,y-1}^{(n)} \\
+\sqrt{(x+1)(y+1)(2j-x)(2j-y)}\rho _{x+1,y+1}^{(n)}\,.
\end{multline}
Note that if the initial state is diagonal in the $|j,m\rangle $ basis (as
is noted in Sec.~\ref{subsec:measurement}, we can demand this without loss
of generality), it remains diagonal. In this case, the recurrence relation
is simply
\begin{align}
  (2j+1)^{2}\rho _{xx}^{(n+1)}
  &=(2j^{2}+2j+1+2(j-x)^{2})\rho_{xx}^{(n)}
  \notag \\
  & \quad +x(2j+1-x)\rho _{x-1,x-1}^{(n)}  \notag \\
  & \quad +(x+1)(2j-x)\rho _{x+1,x+1}^{(n)}\,.  \label{diagrecurrence}
\end{align}

We now explicitly solve for $\rho ^{(n)}$ for the optimal initial
state $ \rho ^{(0)}=|j,j\rangle \langle j,j|$ for $j\gg 1$. The
recurrence relation in this limit is
\begin{multline}
  \rho _{xx}^{(n+1)}=\bigl[1-\tfrac{1}{2j}(2x+1)\bigr]\rho _{xx}^{(n)}
  \label{eq:limitingrecursionrelation} \\
  +\tfrac{1}{2j}\bigl[x\rho _{x-1,x-1}^{(n)}+(x+1)\rho
  _{x+1,x+1}^{(n)}\bigr] \,.
\end{multline}
Expressing the initial condition in matrix form as
$\rho_{00}^{(0)}=1$ and all other elements equal to $0,$ we find the
diagonal matrix elements after $n$ measurements to be
\begin{equation}
  \rho_{xx}^{(n)}=\sum_{k=x}^{n}(-1)^{k+x}\binom{n}{k}\binom{k}{x}\frac{k!}{
  (2j)^{k}}\,,
  \label{DiagMatrixElementsSum}
\end{equation}
which can be expressed in closed form as
Eq.~(\ref{DiagMatrixElementsLimit}).  One can verify that this is
indeed a solution to the recurrence relation.

\section{Optimal measurement of relative phase}

\label{optimalphasemmt}

We seek the optimal measurement for estimating whether two
oscillators are in phase, or $\pi$ out of phase, assuming the second
has a maximum occupation number of $1$ (i.e. the second oscillator
is a qubit), and assuming that the $0$ and $\pi$ relative phases
occur with equal probability.

The total Hilbert space that we need to take account of is
\begin{equation}
  \bigoplus_{n=0}^{\infty}\mathbb{H}_{n}\otimes \left(
  \mathbb{H}_{0}+\mathbb{H} _{1}\right)
  =\bigoplus_{m=0}^{\infty}\mathbb{K}_{m}\,,
\end{equation}
where
\begin{equation*}
  \mathbb{K}_{m}=
  \begin{cases}
  \text{span}\left\{ \left\vert m\right\rangle \left\vert
  0\right\rangle ,\left\vert m{-}1\right\rangle \left\vert
  1\right\rangle \right\} &\text{ for }
  m>0 \\
  \text{span}\left\{ \left\vert 0\right\rangle \left\vert 0\right\rangle
  \right\} &\text{ for }m=0
  \end{cases}
  \,.
\end{equation*}
By Schur's lemma, any positive operator that is invariant under
collective phase rotations must be block diagonal with respect to
$\bigoplus_{m=0}^{\infty}\mathbb{K}_{m}$.  In other words,
\begin{equation}
  E_{\lambda }=\sum_{m=0}^{\infty}\sum_{\mu }w_{m,\mu }^{(\lambda )}G_{m,\mu
  },,
\end{equation}
where $\{G_{m,\mu }\}_{\mu }$ is a POVM on $\mathbb{K}_{m},$ and
where $ w_{m,\mu }^{(\lambda )}\geq 0$ and $\sum_{\lambda }w_{m,\mu
}^{(\lambda )}=1$ to ensure that $\sum_{\lambda }E_{\lambda
}=\mathbb{I}$.  It follows that $ w_{m,\mu }^{(\lambda )}$ is a
probability distribution over $\lambda$ and consequently that the
$E_{\lambda }$ can be obtained by random sampling of the
$G_{m,\mu}$.  Thus, we may as well simply measure the POVM $\left\{
G_{m,\mu }\right\} _{m,\mu }$.

Because the measurement outcome associated with the POVM element
$\mathbb{K}_{0}$ yields no information about the relative phase, we
adopt the arbitrary convention that upon obtaining such an outcome a
relative phase of $0$ is guessed.

Now consider the measurement outcomes associated with POVM elements
confined to $\mathbb{K}_{m}$ for $m>0$.  Within $\mathbb{K}_{m}$,
the optimal POVM for estimating whether the relative phase is $0$ or
$\pi$ may be assumed to be covariant with respect to the group of
relative transformations.  Because there are only two
transformations, namely, $V(0)=I\otimes I$ and $V(\pi )=I\otimes
e^{i \pi \hat{N}},$ there need only be two POVM elements, which we
denote $G_{m,0}$ and $G_{m,\pi },$ and covariance implies that
\begin{equation*}
G_{m,\pi }=V(\pi )G_{m,0}V(\pi )^{\dag }\,.
\end{equation*}

The optimal POVM may also be assumed to be rank 1 in
$\mathbb{K}_{m},$ so that
\begin{equation}
  G_{m,0}=\left\vert g_{m}\right\rangle \left\langle g_{m}\right\vert .
\end{equation}
where
\begin{equation}
  \left\vert g_{m}\right\rangle =g_{m,0}\left\vert m\right\rangle \left\vert
  0\right\rangle +g_{m,1}\left\vert m-1\right\rangle \left\vert 1\right\rangle \,.
\end{equation}
Noting that
\begin{equation}
  V(\pi )\left\vert g_{m}\right\rangle =g_{m,0}\left\vert m\right\rangle
  \left\vert 0\right\rangle -g_{m,1}\left\vert m-1\right\rangle \left\vert
  1\right\rangle \,,
\end{equation}
it follows that
\begin{multline}
  G_{m,0}+G_{m,\pi }=2\left\vert g_{m,0}\right\vert^{2}
  |m\rangle\langle m| \otimes |0\rangle\langle 0|
  \\ +2\left\vert g_{m,1}\right\vert ^{2}|m{-}1\rangle\langle m{-}1|
  \otimes |1\rangle\langle 1| \,.
\end{multline}
But given that $G_{m,0}+G_{m,\pi }=I_{\mathbb{K}_{m}}$, we conclude
that $ \left\vert g_{m,0}\right\vert ^{2}=\left\vert
g_{m,1}\right\vert ^{2}=1/2$. Thus
\begin{equation*}
  \left\vert g_{m}\right\rangle =2^{-1/2}\left( \left\vert m\right\rangle
  \left\vert 0\right\rangle +e^{i\phi _{m}}\left\vert m-1\right\rangle
  \left\vert 1\right\rangle \right)\,,
\end{equation*}
for some phase $\phi _{m}.$  Because we have assumed that the
outcome associated with the POVM element $G_{0}$ is the one which
leads to a guess of $0$ relative phase, the optimal POVM must have
$\phi _{m}=0$.  Thus, $ \left\vert g_{m}\right\rangle =\left\vert
m,+\right\rangle$ and $V(\pi )\left\vert g_{m}\right\rangle
=\left\vert m,-\right\rangle $, where $ \left\vert m,\pm
\right\rangle $ is defined as in Eq.~(\ref{mpm}).

Coarse-graining all the POVM elements that lead to a guess of $0$
relative phase (which includes the projector onto $\left\vert
0\right\rangle \left\vert 0\right\rangle $ according to the above
convention) into a single POVM element $\Pi _{+}$, and all those
that lead to a guess of $\pi $ relative phase into a POVM element
$\Pi _{-}$, we find that the optimal POVM has the form of
Eq.~(\ref{Pi+Pi-}).

\section{Optimal RF state for estimating relative phase}

\label{optimalphasestate}

The characteristic equation we must solve is
\begin{equation}
  \det (M_{N}-\lambda I_{N})=0\,,
\end{equation}
where $I_{N}$ is the identity operator on the space of $N$ or fewer
photons. Defining $G_{N}\equiv M_{N}-\lambda I_{N}$,\ one finds that
\begin{equation}
  \det G_{N}=-\lambda \det G_{N-1}-\det G_{N-2}\,,
\end{equation}
for which the solution is
\begin{equation}
  \det G_{N}=U_{N}(-\lambda /2)\,,
\end{equation}
where the $U_{N}$ are the Chebyshev polynomials of the second kind,
given by $U_{m}(\cos \theta)=\frac{\sin \left[ (m+1)\theta \right]
}{\sin \theta }$.  Given that $U_{N}(x)=\pm U_{N}(-x)$, it follows
that the characteristic equation is $U_{N+1}(\lambda /2)=0$, and
thus the largest eigenvalue is $\lambda^{\max} = 2\cos(\pi/(N+2))$.

To find the eigenvector $|\psi\rangle$ associated with the largest
eigenvalue, we must solve $M_{N}|\psi\rangle =\lambda ^{\max}
|\psi\rangle$.  Defining $\psi_{m}\equiv \langle m|\psi\rangle$, we
have  $\psi_{m+1}+\psi_{m-1}=\lambda^{\max}\psi_{m}$ for $1\leq
m\leq N-1$.  At $m=0$, we have $\psi_{1}=\lambda^{\max}\psi_{0}$,
and at $m=N$, we have $\psi_{N-1}=\lambda^{\max }\psi_{N}$.  The
solution is
\begin{equation}
  \psi_{m} = U_{m}(\lambda^{\max}/2)
  = \mathcal{N} \sin\Bigl[\frac{(m+1)\pi}{N+2}\Bigr] \,,
\end{equation}
where $\mathcal{N}$ is a constant. The coefficients $\psi _{m}$ fall
to zero at $N+1$, verifying the presence of a cut-off in photon
number at $m=N$.  This solution confirms
Eq.~(\ref{eq:optimalphasestate}).

\end{document}